# Architectures and Technologies for a Space Telescope for Solar System Science

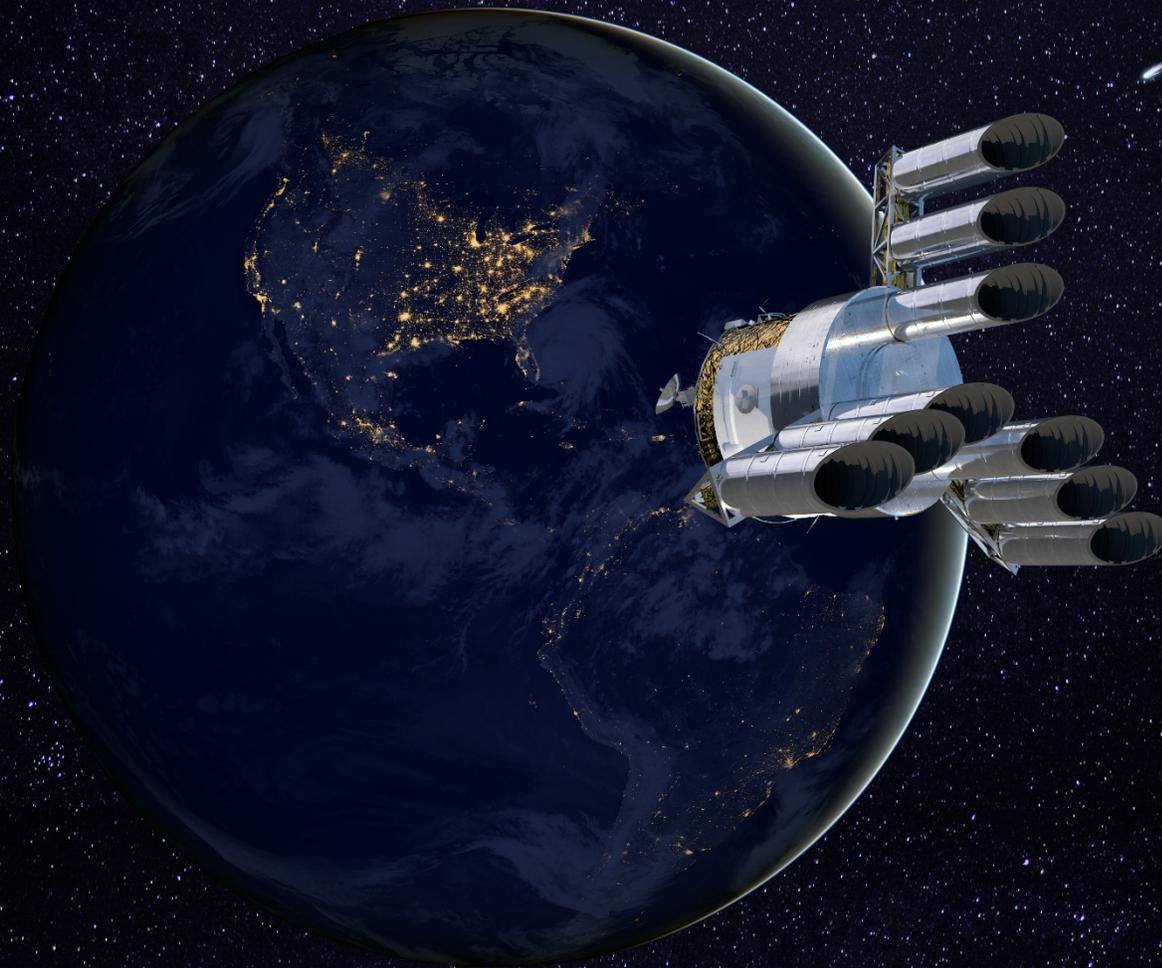

[1]K.M. Sayanagi, [2]C.L. Young, [2]L. Bowman, [3]J. Pitman, [4]B. Naasz, [5]B. Meinke,
[6]T. Becker, [7]J. Bell, [8]R. Cartwright, [9]N. Chanover, [10]J. Clarke, [11]J. Colwell, [12]S. Curry,
[12]I. de Pater, [3]G. Delory, [13]L. Feaga, [14]L.N. Fletcher, [6]T. Greathouse, [15]A. Hendrix,
[16]B.J. Holler, [17]G. Holsclaw, [18]K.L. Jessup, [13]M.S.P. Kelley, [12]R. Lillis, [19]R.M.C. Lopes,
[12]J. Luhmann, [2]D. MacDonnell, [8]F. Marchis, [8]M. McGrath, [4]S. Milam, [20]J. Peralta,
[6]M.J. Poston, [6]K. Retherford, [21]N. Schneider, [12]O. Siegmund, [18]J. Spencer,
[22]R.J. Vervack Jr., [15]F. Vilas, [12]E. Wishnow, [12,8]M.H. Wong

Corresponding Author: Kunio M. Sayanagi; kunio.sayanagi@hamptonu.edu

[1]Hampton University, Hampton, VA
[2]NASA Langley Research Center, Hampton, VA
[3]Heliospace Corporation, Berkeley, CA
[4]NASA Goddard Space Flight Center, Greenbelt, MD
[5]Ball Aerospace, Boulder, CO
[6]Southwest Research Institute, San Antonio, TX
[7]Arizona State University, Tempe, AZ
[8]SETI Institute, Mountain View, CA
[9]New Mexico State University, Las Cruces, NM
[10]Boston University, Boston, MA
[11]University of Central Florida, Orlando, CA
[12]University of California, Berkeley, CA
[13]University of Maryland, College Park, MD
[14]University of Leicester, Leicester, UK
[15]Planetary Science Institute, Tucson, AZ
[16]Space Telescope Science Institute, Baltimore, MD
[17]University of Colorado, Boulder, CO
[18]Southwest Research Institute, Boulder, CO
[19]Jet Propulsion Laboratory, California Institute of Technology, Pasadena, CA
[20]Japan Aerospace Exploration Agency, Sagamihara, Japan
[21]University of Colorado Laboratory for Atmospheric and Space Physics, Boulder, CO
[22]Johns Hopkins Applied Physics Laboratory, Laurel, MD

## 1. Executive Summary

We advocate for a mission concept study for a space telescope dedicated to solar system science in Earth orbit. Such a study was recommended by the Committee on Astrobiology and Planetary Science (CAPS) report "Getting Ready for the Next Planetary Science Decadal Survey." The Mid-Decadal Review also recommended NASA to assess the role and value of space telescopes for planetary science. The need for high-resolution, UV-Visible capabilities is especially acute for planetary science with the impending end of the Hubble Space Telescope (HST); however, NASA has not funded a planetary telescope concept study, and the need to assess its value remains. Here, we present potential design options that should be explored to inform the decadal survey.

Scientific rationale for a space telescope dedicated to solar system science has been presented by a white paper by Young et al[1], who identified a wide range of high-priority scientific investigations that could be enabled by UV-Visible high-resolution observations not achievable with other assets on the ground or in space. Young et al. especially highlighted that a high angular resolution space telescope operating in the UV-Visible range enables observations of temporally dynamic phenomena and compositional surveys of small bodies in the outer solar system. The current white paper identifies potential architectures that should be considered in a concept study to enable a New Frontiers (NF)-class mission that offers UV-Visible high-resolution observation capabilities.

We present two optical designs to illustrate a range of options for a future solar system telescope. The first is a traditional filled-aperture design, and the second is a sparse-aperture telescope. The first option explores a telescope that is fully assembled on the ground and will not require re-configuration in space; consequently, its size is constrained by the launch vehicle. The second is a sparse-aperture telescope with a ~10-meter effective aperture but with a light-collecting area equivalent of a filled ~2-meter aperture; an artist's illustration is presented on the cover page. We advocate for this design trade because a sparse aperture design is uniquely suitable for the high-priority solar system investigations that require the high angular resolution enabled by a >2-meter aperture while the sensitivity requirements can be satisfied by a 2-meter filled-aperture.

The large synthetic aperture of a sparse-aperture telescope will be enabled by deployable mechanisms and/or in-Space Assembly (iSA) technologies; however, how best a sparse aperture telescope can take advantage of these technologies remains to be assessed. For filled-aperture designs, deployable mechanisms are incorporated in the James Webb (JWST). The Astronomy and Astrophysics decadal survey (Astro2020) also examined deployable structures for the Large UV-Optical-Infrared Surveyor (LUVOIR)[2], Habitable Exoplanet Observatory (HabEx)[3] and Origins Space Telescope[4], and the iSA technologies for the iSA Telescope (iSAT)[5]; all four concepts incorporate a filled aperture. However, we note that deployable and iSA technologies have not been systematically compared within a single study. We predict that a sparse-aperture design will uniquely benefit from both and advocate for a new study to examine design options for a future planetary telescope that enables an unprecedented angular resolution within the bounds of the NF Program. Furthermore, a sparse aperture telescope may serve as a technological pathway to enable future large astrophysics space telescopes while serving the needs of the planetary science community.

---

[1] Available at https://bit.ly/2Y2zRFF
[2] https://asd.gsfc.nasa.gov/luvoir/
[3] https://www.jpl.nasa.gov/habex/
[4] https://asd.gsfc.nasa.gov/firs/
[5] https://exoplanets.nasa.gov/exep/technology/in-space-assembly/iSAT_study/

## 2. Overview of Telescope Performance Requirements

Key performance requirements for a space telescope dedicated to solar system observations are:
1. Diffraction-limited imaging with resolution of up to 13 milli-arcsec at 500 nm wavelength
2. Light-collecting area greater than that of a 2-meter filled aperture
3. Imaging and Spectroscopic capability between 100 nm – 2 µm
4. High-cadence, long temporal baseline observations

| Science Questions | Science Objectives | Mission Size Small 1.2 m | Mid./Large 2 m | 10 m |
|---|---|---|---|---|
| Are Venus and Titan volcanically active today? | Search for new evidence of ongoing activity on Venus and Titan | R | R | |
| What drives variability in volcanic and cryovolcanic activity? | Determine the statistics of plume activity | R | R | R |
| What is the composition of magma and cryomagma reservoirs? | Determine composition of lava and surface deposits | R | R | |
| What do the compositions/colors of minor bodies/irregular satellites reveal about planetary migration early in solar system history? | Determine the source population(s) of the Jupiter Trojans and irregular satellites of the giant planets. | D, S | | R |
| What dynamical processes shape minor body populations today? | Determine the source population(s) of the Centaurs. | D, S | | |
| What do the compositions of minor bodies reveal about the radial variations in the solar nebula? | Determine how formation distance influenced KBO surface composition. | D, S | | |
| How does energy/momentum transport vary temporally and spatially in dense atmospheres? | Determine statistics, properties, and evolution of convective events, wave systems, vortices, and jets | R | R | |
| How is atmospheric energy transport modulated by chemical and thermodynamic processes? | Determine the response of horizontal circulation, aerosol properties, and gas composition to internal and solar climate forcing | D | | |
| What is the current outer solar system impactor flux? | Detect and characterize impact ejecta fields in giant planet atmospheres | R, D | | |
| What controls auroral processes on different scales of time and planetary size? | Map auroral emission on terrestrial/gas giant/icy bodies, under varying solar wind and magnetospheric conditions | R | R | R |
| What is the balance between internal/ external control of magnetospheric variability? | Measure the 3D structure and variability of the Io plasma torus at Jupiter and the E-ring at Saturn | | | |
| How do cometary coma and nucleus evolve seasonally or with heliocentric distance ($R_h$)? | Determine coma activity and composition and nucleus reflectance over a range of heliocentric distances | D, S | | |
| What processes dominate in cometary coma? | Determine spatial associations of various coma species, as coma activity and morphology evolves | D, S | | |
| What is the current and past environment of planetary rings across the solar system? | Determine the ring particle size distributions and compositions | R | R | |
| How do ring structures evolve and interact with nearby and embedded moons? | Measure structural profiles and temporal variation | R | R | |

***Table 1.*** *Objectives are partially (yellow) or highly (red) compromised in resolution (R), mission duration (D), or sensitivity (S). UV observations that rely on precision optical alignment are boxed in yellow.*

Table 1 lists the priority science questions to be addressed by a UV-Vis-NIR solar system telescope detailed by Young et al. Using these questions, Table 1 also assesses three example notional optical layouts; the 1.2-meter option represents a filled-aperture that could be realized under the Discovery Program. The 2-meter option represents an upper limit for a filled-aperture ground-assembled telescope, while the 10-meter option represents a sparse-aperture design that must be either deployed or assembled in space. Table 1 also illustrates the challenge of precisely aligning the sub-apertures for imaging in UV with the yellow box surrounding green.

Table 1 also illustrates how a 10-meter telescope will advance planetary science over HST. The HST General Observations (GO) program revolutionized planetary science by providing high-quality snapshots and short movies of various solar system phenomena in UV-Vis. While an HST-like 2-meter telescope satisfies the sensitivity requirement, the next revolutionary advance can be realized by higher resolution, high-cadence observations of these targets to capture temporally dynamic phenomena at a higher frequency over a longer temporal baseline. These assessments assume that a telescope is dedicated to regular-cadence solar system observations for 2 years for the 1.2-meter, and 5 years for the 2- and 10-meter options. These objectives cannot be achieved from ground-based observatories because their resolutions are atmosphere-limited (adaptive optics in visible wavelengths are not expected to be available for the foreseeable future), temporal sampling interferes with atmospheric conditions, and UV wavelengths are not accessible from the ground – these issues will be further exacerbated by the impending end of the HST mission.

## 3. Notional Telescope Concept Examples

Below, we compare the two concepts proposed to the Planetary Mission Concept Studies program: the monolithic 2-meter Planetary Dynamics Explorer (PDX), and the Caroline Herschel High Angular Resolution Imaging & Spectroscopy Multiple Aperture (CHARISMA) telescope.

The trade encompasses the key design characteristics that define the optical throughput and resolution needed to most effectively meet the driving observation needs, as well as how the telescope will be assembled, integrated and tested. The goal is to construct a parametric cost model for the angular resolution as a function of cost for the sparse and filled aperture architectures and find a solution that satisfies the measurement requirements that optimizes the cost.

### 3.1 PDX Concept

The PDX concept's design goal is to maximize the filled aperture size to enable the sensitivity required to address the objectives listed in Table 1 using mature technologies. A recent paper predicts that an HST-class Optical Telescope Assembly (OTA) should be possible in 2016 for approximately $125M ($130M in FY18$), and the OTA accounts for 12% of the total mission cost excluding launch[6], so an HST-class telescope should cost about $1.04B in FY18$.

| Parameters | HST | PDX | Atlas V 400 | Atlas V 500 | Delta IV Medium |
|---|---|---|---|---|---|
| Length | 13.2 m | 11.0 m | 5.8 m (Extra Extended) | 7.6 m (Medium) | 6.5 m |
| Diameter | 4.2 m | 3.5 m | 3.8 m | 4.6 m | 3.8 m |
| Mass | 12,000 kg | 6,900 kg? | 15,718 kg (LEO) 5,860 kg (GTO) | 18,814 kg (LEO) 6,860 kg (GTO) | 13,140 kg (LEO) 4,490 kg (GTO) |

*Table 2. Comparison of HST, PDX, and launch vehicle constraints. PDX scaling assumes a linear factor of 83% HST based on aperture ratio, with an identical focal length.*

| | Sensitivity | Diffraction Limit | Instrument FoV |
|---|---|---|---|
| Measurement Performance | Imaging Limiting Mag = 31 Spec. Limiting Mag = 24 | 63 mas at 500 nm | 110 arcsec 4.2 arcsec/mm Plate Scale |
| Baseline Design | 2-meter circular aperture | 2-meter circular aperture | 58 m Focal Length |

*Table 3. PDX baseline performance*

The PDX concept adopts a Ritchey-Chretien (RC) optical design like HST. Launching a 2-meter RC is a challenge, as telescope length is large compared to typical fairings (Table 2). A preliminary analysis provides the notional performance listed in Table 3, assuming a 2Kx2K sensor array (10 µm pitch) over a 110" FOV. A detailed study is needed to reduce length (and spacecraft mass/volume). Mass reduction is also needed to reach a high orbit; Table 2 shows the mass capacity of the launch vehicles to a Geostationary Transfer Orbit (GTO) to illustrate the challenge.

Although a 2-meter filled aperture telescope does not represent an advance over HST in terms of sensitivity and angular resolution, we advocate for a study to assess its cost to serve as a reference point to understand the value of an advanced CHARISMA concept. With HST-like capabilities, PDX can address high-priority objectives through dedicated planetary observations that enables high-cadence long-duration temporal surveys of time-variable phenomena.

---

[6] Stahl and Henrichs, in *Modeling, Systems Engineering, and Project Management for Astronomy VI*, 2016.

## 3.2 CHARISMA Concept

The design goal of CHARISMA is to innovatively revolutionize the angular resolution within the constraints of NF. To enable a large effective aperture, the CHARISMA telescope gathers light using multiple Cassegrain sub-apertures, which together enable diffraction-limited resolution of the combined aperture radius. This sparse-aperture design maximizes angular resolution without the massive structure necessary for a filled-aperture telescope. The cover graphic and Fig. 1 illustrate a concept with nine 1-meter sub-apertures in a tri-arm configuration to form a 10-m effective aperture. The collimated beams emerging from the sub-apertures are brought in phase by the optical path adjustment mechanisms in the arm trusses. Once the beams are in phase, the combiner optics focus the beams on a single focal plane. The notional design has a light-collecting area equivalent of HST and a diffraction limit of a 10-m telescope. Chamfered baffles on the sub-apertures reduce the exclusion zone to about 30º around the sun, facilitating observations of Venus and comets near perihelion, and the outer planets when they are near solar conjunction.

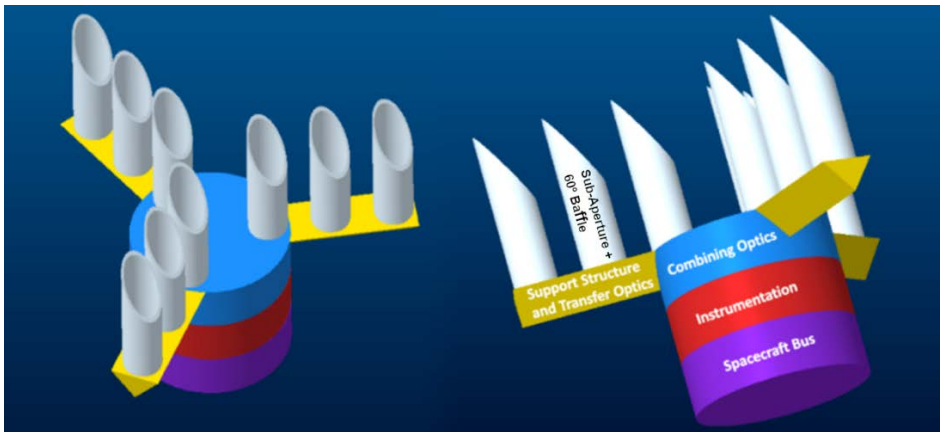

*Fig. 1. CHARISMA in fully deployed configuration.*

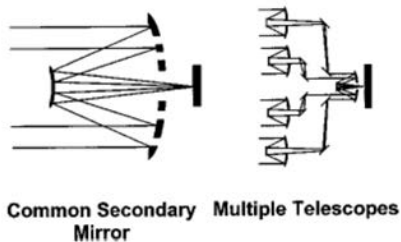

*Fig. 2. Sparse-aperture telescope design options: Common Secondary Mirror, and Multiple Telescopes[7]*

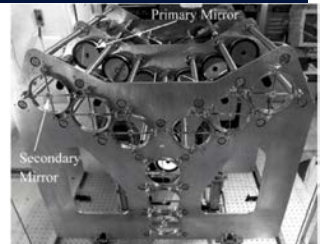

*Fig 3. Tri-Arm sparse aperture telescope prototype[8].*

A sparse-aperture telescope design either has a common secondary mirror (CSM), or multiple telescopes (MT) as shown in Fig. 2. We adopt the MT approach primarily because CSM is incompatible with the chamfered baffles around the sub-apertures. The MT architecture of CHARISMA is similar to the MIDAS concept designed for the Jupiter Icy Moon Orbiter, whose heritage traces to Lockheed-Martin Advanced Technology Center's work on distributed aperture telescopes; Fig. 3 shows a Lockheed-Martin Tri-Arm prototype[8]. A detailed cost study is needed to determine the optimal architecture enabled within NF cost.

The main design options that affect the telescope performance are the size/number of the sub-apertures, and the fill factor (the fraction of light collecting area within the effective aperture). Sub-aperture geometric configurations should also be studied; the sub-aperture layout tunes the Fourier sampling in the spatial dimensions and affects the shape of the modulation transfer function (MTF; Fig. 4). The cover graphic and Fig. 1 depict a Tri-Arm design; alternate configurations such as Golay-6 (Fig 4-left) should also be studied. The MTF of the optical train should be modeled in the entire wavelength range to minimize artifacts of the sub-aperture layout.

---

[7] Fiete et al. 2002, *Optical Engineering,* vol. 41, pp. 1957-1969.
[8] Pitman et al. *in Instruments, Science, and Methods for Geospace and Planetary Remote Sensing*, 2004.

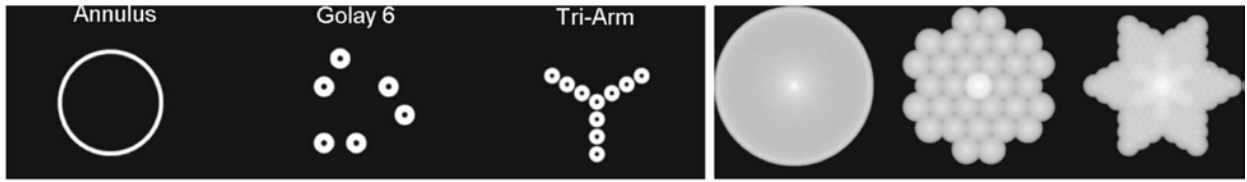

*Fig. 4 Left*: Annular, Golay-6, and Tri-Arm sub-aperture layouts with 16% fill factor.
*Right*: Modulation Transfer Functions corresponding to the sub-aperture arrangements shown in left[7].

The length and alignment of beam paths must be actively adjusted for both coarse (mm) and fine (nm) misalignments, which would pose significant challenges in the UV wavelengths. Our preliminary assessment determined that the 10-nm precision required to phase 100 nm wavelength beams is realistic, based on the 7.5-nm resolution TRL 7 actuators developed for JWST mirrors[9].

For a preliminary design with nine 1-m sub-apertures and four instruments to be assembled by one robotic arm in low-Earth orbit, NASA Langley Research Center Basis-of-Estimate tool predicts that CHARISMA will cost $815M including 30% reserve, in which the OTA and instruments account for 23% and 37% of the cost. In contrast, a typical space telescope cost includes 12% OTA and 25% instruments[6]; the high OTA cost for CHARISMA reflects its complexity; nevertheless, a preliminary estimate suggests that CHARISMA could be within the $1B NF cost cap.

## 4. Telescope Deployment and Assembly Technologies

While the PDX concept aims for a high-TRL design, CHARISMA requires deployable and/or iSA technologies to overcome launch constraints. This section reviews the recent space telescope designs that incorporate deployable structures and iSA technologies to illustrate that the benefits of these technologies have not been examined for a sparse aperture telescope like CHARISMA.

### 4.1. Deployable Structures: Lessons from Recent Studies

The current development in deployable structure is evidenced in ongoing projects such as JWST and the Imaging X-ray Polarimetry Explorer (IXPE). JWST incorporates a folded mirror and sunshield that will unfurl in space while IXPE deploys a 5.5-m boom to realize the long focal length of the telescope. As IXPE is in Earth orbit and experiences thermal cycling, special considerations are needed for the boom to ensure that its length remains stable. Many of these deployment technologies should be applicable to a sparse-aperture telescope; however, their applicability is yet to be examined. JWST offers lessons about how the large number of complex deployable mechanisms has impacted the cost and schedule to assure reliability in the Verification and Validation (V&V) of those structures on the ground. Another JWST lesson is that a deployable structure's joints and actuators must be built to withstand the launch load, even though most of the moving parts are needed only for one-time deployment in zero-G; the requirement for all the mechanisms to withstand launch has proven to be a significant impact on the cost.

A notable factor that has precluded a multi-aperture design from being considered in astrophysics missions is the desire to detect and characterize exoplanets using a coronagraph. Because a coronagraph places an obstruction in the aperture to block the star to observe nearby planets, it is fundamentally incompatible with a distributed aperture telescope. In addition, because astrophysics missions pursue higher sensitivities to observe fainter targets, sparse aperture systems do not align with their objectives. Thus, astrophysics mission studies have not favored multi-aperture systems, and all four of the Astro2020 telescope concepts have a large filled aperture with a coronagraph, of which three (LUVOIR, HabEx and Origins) have deployable structures.

---

[9] Stahl, 2009: https://ntrs.nasa.gov/search.jsp?R=20090028784

An alternative approach to observe an exoplanet is to use an interferometer in which light waves from multiple apertures are combined such that the signal from the star is destructively interfered and cancelled while the planets' signals are constructively interfered – this technique is incorporated in the Terrestrial Planet Finder Interferometer (TPF-I) and the Space Interferometery Mission (SIM) concepts, both of which have multiple sub-apertures but neither concept has advanced due to budgetary constraints. How deployable structures can benefit a sparse-aperture space telescope is better studied first in the context of advancing solar system science objectives because solar system observations do not require a coronagraph, and the size of the sub-apertures can be much smaller for solar system observations than for interferometers targeting exoplanets.

### 4.2. in-Space Assembly Technologies: Lessons from Recent Studies

iSA technologies also enable large systems in space; a key difference between deployable structure and iSA is that, while a deployable structure transforms from its launch configuration to an operational configuration by unfurling numerous joints controlled by actuators, the iSA technique concentrates all actuators on a single/few Robotic Servicing Arm(s) (RSA)[10], and modular components are rigidly mounted in a configuration optimized to withstand launch load.

A recent Astro2020 study of the iSAT concept offers some guidance for a planetary telescope. In particular, the iSAT study determined that the iSA technologies become enabling for a space telescope that has a primary mirror diameter of 15-meter or greater and may reduce the cost of smaller telescopes[11]. However, even though iSAT's primary mirror is segmented, its aperture is filled and it essentially employs the CSM design illustrated in Fig. 2. We hypothesize that the cost function derived for a filled CSM design is not applicable to a sparse MT-architecture like CHARISMA. A sparse aperture design is beneficial specifically to planetary science as the observations are primarily limited by angular resolution rather than sensitivity; thus, we advocate for a concept study of a sparse aperture telescope as part of the planetary science decadal survey.

Furthermore, recent developments make examining iSA technologies for space telescopes timely. In 2018, NASA, other agencies, and commercial industry formed a nationally coordinated effort, hosted by the NASA Office of the Chief Technologist Science and Technology Partnership Forum to pursue Orbital Servicing, Assembly and Manufacturing (OSAM) at a national level[12, 13]. CHARISMA was initially conceived to leverage technologies under OSAM including RSA, precision couplers, grapple features, and vision systems required in iSA. The OSAM effort also examines approaches to component modularization, module packaging, module launch integration and module assembly, which are also beneficial to advance deployable structures.

### 4.3. Deployable and in-Space Assembly Technologies: Common Challenges

We recommend a study to identify where the deployable and iSA technologies' applicability diverges, which is key to building a roadmap to a telescope that has the right design and size to be cost-effective and low-risk in the coming decade. A study is needed in particular to examine the technology areas relevant to both deployable and iSA to test how the approaches could be combined, including (1) Modular Concepts, (2) Mechanisms, (3) Software, and (4) V&V.

---

[10] https://sspd.gsfc.nasa.gov/robotic_servicing_arm.htm
[11] https://exoplanets.nasa.gov/exep/technology/in-space-assembly/iSAT_study/
[12] Arney etal. 2018 in *2018 AIAA SPACE and Astronautics Forum and Exposition*.
[13] https://www.nasa.gov/sites/default/files/atoms/files/st_partnership_isa_open_forum_presentations.pdf

In particular, V&V is expected to be a significant challenge. Because few large space observatories have flight heritage, CHARISMA would face the same challenges as JWST, IXPE and other advanced astrophysics telescope concepts that must design customized in-space V&V processes. Prior to any hardware build, CHARISMA would be broken down into modules through trade analysis to determine which components, subsystems, and locations are best suited for modularization. During this phase, module interfaces would be identified as part of understanding how to assemble/deploy the system using currently available mechanisms. System stability data and performance requirements based on the heritage of space observatories such as HST, JWST and the Roman Space Telescope are available in the analyses. V&V may be feasible on the ground for some components while for others it must be done in-space. A study should include cost/risk trades to determine what level of V&V is needed to adequately demonstrate flight readiness.

## 5. Connection to Astrophysics Telescope Studies

A 10-meter class sparse-aperture planetary telescope can serve as a pathfinder for larger astrophysics missions. The iSAT report recommended that modularization and V&V technologies be matured in the 2020s to construct a 20-meter telescope in the 2030s to be launched in the 2040s.

The final reports of the four Astro2020 telescope studies (LUVOIR, HabEx, Origins and iSAT) uniformly identify the need to advance the deployment/assembly technologies as well as the metrology and V&V approaches. The CHARISMA concept offers an opportunity to demonstrate several aspects of these least mature technologies in the early to mid-2020s by using its modularized telescope design with far less demanding requirements than those of an astrophysics telescope. Specifically, CHARISMA will offer an opportunity to fly and mature the following technologies to TRL 9: (1) Deployment/Assembly mechanisms such as precision latches and bolts that enable large structures and yet stable in the presence of on-orbit disturbances, and (2) Advanced metrology methods for in-space pre-assembly inspection and post assembly V&V. Thus, CHARISMA will be complementary and beneficial to future astrophysics missions.

In addition, many challenges encountered by a space-based optical interferometer are analogous to those of sparse aperture telescopes. Perhaps a sparse-aperture telescope targeting solar system science could serve as a technology demonstrator for a future astrophysics interferometric mission; such a possibility should be considered in the concept study advocated here.

## 6. Recommendation

We recommend a study to understand how deployable and iSA technologies enable a large sparse aperture space telescope optimized for solar system observations and understand their benefits over a traditional ground-assembled, filled-aperture telescope within the constraints of the NF program. Such a study is needed because while studies to date have examined deployable and iSA technologies separately for space telescopes, no study has compared the benefits, development timeline, cost, or next steps in an apples-to-apples comparison of the two. A new study should also systematically examine the science objectives enabled by the two approaches.

A sparse aperture space telescope is uniquely advantageous in satisfying solar system observation requirements. Thus, the planetary science decadal survey is an appropriate opportunity to conduct a systematic comparison of deployable and iSA technologies for a sparse aperture telescope. This whitepaper also demonstrates that a funded pre-formulation activity that thoroughly conducts the deployable vs. iSA architecture trade is the essential next step towards a future large space telescope. A dedicated planetary telescope is the perfect opportunity to take this step, which also serves as an important technology pathfinder for future astrophysics observatories with much larger apertures and more stringent alignment and stability requirements.